\documentclass[article, shortnames, nojss]{jss}\usepackage[]{graphicx}\usepackage[]{color}
\makeatletter
\def\maxwidth{ %
  \ifdim\Gin@nat@width>\linewidth
    \linewidth
  \else
    \Gin@nat@width
  \fi
}
\makeatother

\usepackage{Sweave}

\usepackage{amsmath}
\usepackage{amssymb}
\usepackage{algorithm}
\usepackage{algorithmic}
\usepackage{tabulary}
\usepackage[usenames,dvipsnames]{xcolor}
\usepackage{hyperref}

\newcommand{\bgamma}{\boldsymbol{\gamma}}
\newcommand{\btheta}{\boldsymbol{\theta}}

\newcommand{\by}{\boldsymbol{y}}
\newcommand{\bd}{\boldsymbol{d}}
\newcommand{\bx}{\boldsymbol{x}}
\newcommand{\argmax}{\mathrm{arg~max}}

\newcommand{\bSigma}{\boldsymbol{\Sigma}}

\newcommand{\rev}[1]{{\color{black} #1}}

\author{Antony M. Overstall \\ University of Southampton \And 
David C. Woods \\ University of Southampton \And 
Maria Adamou\\ University of Southampton}
\title{\pkg{acebayes}: An \proglang{R} Package for Bayesian Optimal Design of Experiments via Approximate Coordinate Exchange}

\Plainauthor{Antony M. Overstall, David C. Woods, Maria Adamou} 
\Plaintitle{acebayes: An R Package for Bayesian Optimal Design of Experiments via Approximate Coordinate Exchange} 
\Shorttitle{\pkg{acebayes}: Bayesian Design via Approximate Coordinate Exchange}

\Abstract{
We describe the \proglang{R} package \pkg{acebayes} and demonstrate its use to find Bayesian optimal experimental designs. A decision-theoretic approach is adopted, with the optimal design maximising an expected utility. Finding Bayesian optimal designs for realistic problems is challenging, as the expected utility is typically intractable and the design space may be high-dimensional. The package implements the approximate coordinate exchange algorithm to optimise (an approximation to) the expected utility via a sequence of conditional one-dimensional optimisation steps. At each step, a Gaussian process regression model is used to approximate, and subsequently optimise, the expected utility as the function of a single design coordinate (the value taken by one controllable variable for one run of the experiment).  In addition to functions for bespoke design problems with user-defined utility functions, \pkg{acebayes} provides functions tailored to finding designs for common generalised linear and nonlinear models. The package provides a step-change in the complexity of problems that can be addressed, enabling designs to be found for much larger numbers of variables and runs than previously possible. We provide tutorials on the application of the methodology for four illustrative examples of varying complexity where designs are found for the goals of parameter estimation, model selection and prediction. These examples demonstrate previously unseen functionality of \pkg{acebayes}.    
}


\Address{
  Antony M. Overstall, David C. Woods and Maria Adamou\\
  Southampton Statistical Sciences Research Institute\\
  University of Southampton\\
  Southampton, SO17 1BJ, UK\\
  E-mail: \email{\{A.M.Overstall,D.Woods,M.Adamou\}@southampton.ac.uk}\\
}
\IfFileExists{upquote.sty}{\usepackage{upquote}}{}
\begin{document}

\newpage
\section[Introduction]{Introduction}

A well-planned and executed experiment is an efficient and effective way of learning the effect of an intervention on a process or system \citep{BHH}, and design of experiments is a key contribution of statistics to the scientific method (\citealp{Stigler}, ch.~6). Statistical design is an ``a priori'' activity, taking place before data is collected, and so fits naturally within a Bayesian framework. Before experimentation, current knowledge on models and parameters can be represented by prior probability distributions, and the experimental aim (e.g., parameter estimation, model selection, or prediction) can be incorporated into a decision-theoretic approach through the specification of a utility function (\citealp{Berger1985}, ch.~2). A Bayesian optimal design is then found by maximising the expectation of this utility over the space of all possible designs, where expectation is with respect to the joint distribution of all unknown quantities including the, as yet, unobserved responses \citep{ChalonerVerdinelli}.

To formalise, suppose the aim of an experiment that varies $k$ variables in $n$ runs is to estimate or identify quantities $\bgamma = (\gamma_1,\ldots,\gamma_p)^\top$, which, for example, may be (functions of) parameters, model indicators or future responses. We perform this task using data $\by = (y_1,\ldots,y_n)^\top\in\mathcal{Y}\subset\mathbb{R}^n$ collected using a design $\bd\in\mathcal{D}\subset\mathbb{R}^{n\times k}$, an $n\times k$ matrix with $i$th row $\bx_i^\top = (x_{i1},\ldots,x_{ik})$ holding the treatment, or combination of values of the controllable variables, assigned to the $i$th run of the experiment ($i=1,\ldots,n$). We refer to $nk$ as the dimensionality of the design and the $x_{ij}$ as \textit{coordinates} of the design ($j=1,\ldots,k$). 

A decision-theoretic Bayesian optimal design $\bd^\star$ maximises the expected utility
\begin{eqnarray}
U(\bd) & = & \mathrm{E}_{\bgamma,\by | \bd} \left[u(\bgamma,\by,\bd)\right] \nonumber \\
& = & \int u(\bgamma,\by,\bd) \pi(\bgamma,\by | \bd)\, \mathrm{d}\bgamma \,\mathrm{d}\by \label{eqn:exputil} \\
& = & \int u(\bgamma,\by,\bd) \pi(\bgamma | \by, \bd)\pi(\by | \bd)\, \mathrm{d}\bgamma \,\mathrm{d}\by \label{eqn:exputil1} \\
& = & \int u(\bgamma,\by,\bd) \pi(\by | \bgamma, \bd)\pi(\bgamma | \bd)\, \mathrm{d}\bgamma \,\mathrm{d}\by\,, \label{eqn:exputil2}
\end{eqnarray}
with utility function $u(\bgamma,\by,\bd)$ providing a measure of success of design $\bd$ for quantities $\bgamma$ and data $\by$, and $\pi(\cdot | \cdot)$ denoting a conditional probability density or mass function. The density/mass $\pi(\bgamma | \bd)$ quantifies prior information about $\bgamma$ available before the experiment is run. The equivalence of Equations~\ref{eqn:exputil1} and~\ref{eqn:exputil2} follows from application of Bayes theorem; Equation~\ref{eqn:exputil1} more clearly shows the dependency on the posterior distribution, whereas Equation~\ref{eqn:exputil2} is often more useful for calculations and computation.

Although straightforward in principle, there are several hurdles to the practical evaluation and optimisation of Equations~\ref{eqn:exputil1} or~\ref{eqn:exputil2}, highlighted in recent reviews by \citet{ryan_drovandi_mcgree_pettitt_2015} and \citet{woods_ace_2016}.
\begin{enumerate}
\item The expected utility often results from an analytically intractable and, typically, high-dimensional integral. For such utilities, $U(\mathbf{d})$ may be approximated by a weighted sum
\begin{equation}\label{eqn:approxEU}
\tilde{U}(\bd) = \sum_{b=1}^B w_b u(\bgamma_b, \by_b, \bd)\,,
\end{equation}
for $w_b>0$. For example, a Monte Carlo approximation would sample $\{\bgamma_b, \by_b\}_{b=1}^B$ from the joint distribution $\pi(\bgamma,\by | \bd)$ and set $w_b = 1/B$; maximisation of $\tilde{U}(\mathbf{d})$ would be a stochastic optimisation problem. For simple utility functions not depending on data $\by$, a deterministic quadrature approximation may be applied, with $\bgamma_b$ and $w_b$ being quadrature abscissae and weights, respectively.
\item The design space may be continuous and of high dimension.
\item The utility function itself may not be available in closed form, requiring an approximation $\tilde{u}(\bgamma,\by,\bd)$ to be substituted into Equation~\ref{eqn:approxEU}. 
\end{enumerate}

Most Bayesian optimal design methodology in the literature has been restricted to low-dimensional designs, i.e., small values of $nk$. See \citet{MullerParmigiani}, \citet{Muller1999}, \citet{amzal_bois_parent_robert_2006}, \citet{Laplace}, and \citet{drovandi_mcgree_pettitt_2013,drovandi_mcgree_pettitt_2014}, where the largest designs found had $nk = 4$. To find designs with larger $n$ for $k=1$ variable, \citet{ryan_drovandi_thompson_pettitt_2014} chose design points as quantiles of a parametric distribution. Although such an approach reduces the dimension of the optimisation problem (e.g., to finding the optimal values of a small number of parameters controlling the distribution), the original optimal design problem is not being directly addressed and hence usually sub-optimal designs will be found.

\citet{woods_overstall_2016} recently presented the first general methodology for finding high-dimensional Bayesian optimal designs using the approximate coordinate exchange (ACE) algorithm. As will be described in Section~\ref{sec:ace}, the main feature of this approach is the combination of the low-dimensional smoothing methodology of \citet{MullerParmigiani} with a coordinate exchange, or cyclic ascent, algorithm (\citealp{MeyerNachtsheim}; \citealp[][p.~171]{lange_2013}). In essence, the high-dimensional, computationally expensive and often stochastic optimisation problem is reduced to a sequence of one-dimensional, computationally cheaper, and deterministic optimisations. In this paper, we describe the \proglang{R} package \pkg{acebayes} \citep{acebayes} which implements the ACE algorithm, and introduce functionality that facilitates finding optimal designs for common classes of models, including nonlinear and generalised linear models. The package provides the first general-purpose software for finding fully-Bayesian optimal designs. In addition, the package implements methods to find pseudo-Bayesian designs (see Section~\ref{sec:approxutil}) using coordinate exchange and quadrature approximations (see \citealp{gotwalt_jones_steinberg_2009}). The package has been demonstrated by finding Bayesian optimal designs for non-trivial statistical models, addressing design spaces with dimensionality approaching two orders of magnitude greater than existing methods.

There are only a few other \proglang{R} packages that attempt to find optimal designs, and none that tackle the general Bayesian design problem addressed by \pkg{acebayes}. The package \pkg{AlgDesign} \citep{AlgDesign} implements exchange-type algorithms to find $D$-, $A$- and $I$-optimal designs for linear models. The package \pkg{OptimalDesign} \citep{OptimalDesign} tackles similar linear model design problems using various algorithms including integer quadratic programming. For nonlinear models, package \pkg{ICAOD} \citep{ICAOD} can be used to find designs for quite general classes of models under ``optimum-on-average'' criteria, amongst others. Such criteria are mathematically equivalent to ``pseudo-Bayesian'' design criteria, see Section~\ref{sec:utility}. This package uses the meta-heuristic imperialist competitive algorithm \citep{EHW2016}. For special classes of nonlinear models, locally optimal designs, with a point mass prior for $\bgamma$ and utility functions not depending on $\by$, can be found by packages \pkg{LDOD} \citep{LDOD}, \pkg{designGLMM} \citep{designGLMM} and \pkg{PopED} \citep{PopED}. 

This paper is structured as follows. We briefly describe both the ACE algorithm and its implementation in \pkg{acebayes} in Section~\ref{sec:ace_acebayes}. Section~\ref{sec:utility} presents common utility functions employed in Bayesian design, and discusses their computational approximation. In Section~\ref{sec:examples} we demonstrate the use of various functions in the \pkg{acebayes} package to find optimal Bayesian designs for common nonlinear and generalised linear models, and bespoke model selection and prediction problems. We conclude in Section~\ref{sec:disc} with a short discussion.   

\section[ACE]{Approximate coordinate exchange and \pkg{acebayes}}
\label{sec:ace_acebayes}


In this section we give a brief description of the ACE algorithm. Full details of the methodology can be found in \citet{woods_overstall_2016}. 

The algorithm has two phases, both of which are provided in full in Appendix~\ref{app:ace}. In Phase I a smooth, and computationally inexpensive \textit{emulator} for the approximation $\tilde{U}(\mathbf{d})$ in Equation~\ref{eqn:approxEU} is maximised as a function of each design coordinate $x_{ij}$ in turn, conditional on the values of the other $nk-1$ coordinates. In essence, the optimisation problem is solved via a sequence of computer experiments (see \citealp{SWN2003}). 

For a stochastic approximation to the expected utility (e.g., Monte Carlo integration), the coordinate value that maximises each emulator is accepted with probability obtained from a Bayesian test of equality of the approximations under the proposed and current designs. For a deterministic approximation (e.g., quadrature), the design proposed by the emulator is accepted if the value of $\tilde{U}(\mathbf{d})$ for the proposed design is larger than for the current design. 

As Phase I tends to produce clusters of design points, Phase II of the algorithm can be applied to attempt to amalgamate these clusters through use of a point exchange algorithm with a candidate set formed from the points in the final Phase I design.       


\subsection[ACE algorithm]{ACE algorithm}
\label{sec:ace}

Phase I of the algorithm (Appendix~\ref{alg:ace1}) uses cyclic ascent to maximise approximation $\tilde{U}(\mathbf{d})$ to the expected utility. A one-dimensional emulator of $\tilde{U}(\mathbf{d})$ is built for each coordinate $x_{ij}$ ($i=1,\ldots,n;\,j=1,\ldots,k$) in turn as the mean of the posterior predictive distribution conditioned on a small number of evaluations of $\tilde{U}(\mathbf{d})$ and assuming a Gaussian process (GP) prior (see \citealp{RW2006}, ch.~2). For the $ij$th coordinate, an emulator is constructed by (i) selecting a one-dimensional space-filling design, $x^1_{ij},\ldots,x^Q_{ij}$, with $Q$ points; (ii) constructing the $Q$ designs $\bd_{ij}^q$, with the $q$th design having $i$th run $\bx_i^q = (x_{i1},\ldots,x_{ij-1}, x_{ij}^q, x_{ij+1},\ldots,x_{ik})^\top$ and all other runs equal to those from the current design; (iii) evaluating $\tilde{U}(\bd_{ij}^q)$ for $q=1,\ldots,Q$; and (iv) fitting a GP regression model to the ``data'' $\{x_{ij}^q, \tilde{U}(\bd_{ij}^q)\}_{q=1}^Q$ and constructing an emulator $\hat{U}_{ij}(x)$ as the mean of the posterior predictive distribution. Maximisation of $\hat{U}_{ij}(x)$ to obtain $x_{ij}^\star$ is via evaluation of the emulator for a large discrete grid of values of $x_{ij}$ to produce design $\bd_{ij}^\star$ with $i$th row $(\bx_i^\star)^\top = (x_{i1},\ldots,x_{ij-1}, x_{ij}^\star, x_{ij+1},\ldots,x_{ik})$. \citet{woods_overstall_2016} found this approach to maximising $\hat{U}_{ij}(x)$ to be robust to multi-modal emulators and computationally efficient due to the negligible computational expense of evaluating the predictive mean.   

%

If $\tilde{U}(\bd)$ is a Monte Carlo approximation, it is subject to two sources of potential errors: Monte Carlo error and emulator inadequacy. To separate these components and reduce the impact of a poor emulator, $\bd_{ij}^\star$ is only accepted as the next design in the algorithm with probability $p^\star$ obtained from a Bayesian hypothesis test, independent of the GP emulator (see Step \ref{alg:1_2e} in Appendix \ref{alg:ace1}). Here $p^\star$ is calculated as the posterior probability that the expected utility for the proposed design $\bd_{ij}^\star$ is greater than that for the current design, given independent Monte Carlo samples of the utility under each design and assuming normally distributed utility values. For cases where this latter assumption is violated, \citet{overstall_etal_2016} developed an alternative procedure derived from a one-sided test of a difference in proportions appropriate for use with, for example, a 0-1 utility function (see Equation~\ref{eqn:01} in Section~\ref{sec:cuf}). Larger Monte Carlo sample sizes are typically used for these tests than for the construction of the emulator to increase the precision of approximation $\tilde{U}(\mathbf{d})$. Both tests are implemented in \pkg{acebayes}. 

For a deterministic $\tilde{U}(\bd)$, design $\bd_{ij}^\star$ is accepted if its approximate expected utility, evaluated independently of the emulator, is greater than that of the current design.  

Phase I can produce clusters of design points where, for example, design points $\bx_i$ and $\bx_{i^\prime}$ are separated by only a small Euclidean distance for some $i,i^\prime = 1,\dots,n$. Often, the design can be improved by consolidating such points into a single repeated design point (see also \citealp{gotwalt_jones_steinberg_2009}). Phase II of ACE (Appendix~\ref{alg:ace2}) performs this consolidation step using a point exchange algorithm \citep[e.g.,][ch.~12]{atkinson_donev_tobias_2007} with a candidate set given by the final design from Phase I. For Monte Carlo approximations to the expected utility, comparison of the approximate expected utility between two designs is again made on the basis of a Bayesian hypothesis test (see Step~\ref{alg:2_2e} of Appendix~\ref{alg:ace2}).

In both phases, convergence is assessed informally using trace plots of the evaluations of approximate expected utility at each iteration. See Section~\ref{sec:logreg} for an example of such a plot produced by \pkg{acebayes}.

Similar to all coordinate exchange algorithms \citep[e.g.,][pg.~36]{goos_jones_2011}, ACE can be sensitive to the starting design. For this reason, it should be repeated from $C$ different starting designs and \code{acebayes} provides methods to facilitate this repetition, potentially via simple parallel computing.

\subsection[acebayes implementation of ACE]{\pkg{acebayes} implementation of ACE} \label{sec:acebayes}

The main functions in the \pkg{acebayes} package are \code{ace} and \code{pace}, which implement both phases of the ACE algorithm and have mandatory and optional arguments as given in Tables~\ref{tab:acearg1} and~\ref{tab:acearg2}, respectively. The \code{ace} function implements the ACE algorithm from a single starting design, whereas \code{pace} repeats ACE from $C$ different starting designs. The argument \code{utility} gives the user complete flexibility to specify the design problem including the choice of statistical model, prior distribution, experimental aim and any necessary approximation to the utility function (see Section~\ref{sec:utility}). 

Much of the \pkg{acebayes} codebase is written in \proglang{C++} and makes use of packages \pkg{Rcpp} \citep{Rcpp} and \pkg{RcppArmadillo} \citep{RcppArmadillo}. Space-filling designs to build the one-dimensional GP emulators are found using the \proglang{R} package \pkg{lhs} \citep{lhs_pack} which generates Latin hypercube samples. 

\begin{table}
\begin{center}
\begin{tabulary}{\textwidth}{lL} \hline
\textbf{Argument} & \textbf{Description} \\ \hline
\code{utility} & A function with two arguments: \code{d} and \code{B}. \\
& For a Monte Carlo approximation (\code{deterministic = FALSE}), it should return a \textbf{vector} of length $B$ where each element gives an evaluation of the (approximate) utility function $\tilde{u}(\bgamma_b, \by_b,\bd)$ for design \code{d} for each pair $\left(\bgamma_b,\by_b\right)$ generated from the joint distribution of $\bgamma$ and $\by$ for $b= 1,\ldots,B$. \\
& For a deterministic approximation (\code{deterministic = TRUE}), it should return a \textbf{scalar} giving the approximate value of the expected utility for design \code{d}. In this latter case, the argument \code{B} can be a list containing tuning parameters for the deterministic approximation. If \code{B} is not required, the utility function must still accept the argument, e.g., using the \code{\ldots} notation.
\\ \hline
\code{start.d} & For \code{ace}: an $n \times k$ matrix specifying the starting design for Phase I (see Step \ref{alg:start} in Appendix \ref{alg:ace1}).\\ 
& For \code{pace}: a list of $C$ different starting designs. \\ \hline
\end{tabulary}
\end{center}
\caption{Mandatory arguments to the \code{ace} and \code{pace} functions. \label{tab:acearg1} }
\end{table}

\begin{table}
\begin{center}
\begin{tabulary}{\textwidth}{lL} \hline
\textbf{Argument} & \textbf{Description} \\ \hline
\code{B} & For a Monte Carlo approximation (\code{deterministic = FALSE}), a vector of length two specifying the size of the Monte Carlo samples generated from the joint distribution of unknown quantities and unobserved responses, to use when approximating the expected utility via Equation~\ref{eqn:approxEU}. The first element specifies the sample size to use in the comparison procedures (see Steps~\ref{alg:1_2e} and~\ref{alg:2_2e} in Appendices~\ref{alg:ace1} and~\ref{alg:ace2}, respectively). The second element specifies the sample size to use for the evaluations of Monte Carlo integration that are used to fit the Gaussian process emulator (see Step~\ref{alg:approx} in Appendix~\ref{alg:ace1}). If missing when \code{deterministic} \code{= FALSE}, the default value is \code{c(20000,1000)}. \\
& For a deterministic approximation (\code{deterministic = TRUE}), \code{B} may be a list of length two containing any necessary tuning parameters for the utility calculations for the comparison and emulation steps.\\ \hline
\code{Q} & The number, $Q$, of evaluations of the approximation to the expected utility function (\ref{eqn:exputil}) used to construct the GP emulator. The default is \code{Q = 20}.\\ \hline
\code{N1} & The number, $N_1$, of iterations of Phase I. The default is \code{N1 = 20}.\\ \hline
\code{N2} & The number, $N_2$, of iterations of Phase II. The default is \code{N2 = 100}.\\ \hline
\code{lower} & Lower limits on the design space. It can be a scalar (all elements have the same lower limit) or an $n \times k$ matrix so that all elements can have unique lower limits. The default is \code{lower = -1}.\\ \hline
\code{upper} & Upper limits on the design space; see \code{lower}. The default is \code{upper = 1}. \\ \hline
\code{limits} & A function that can be used to define complex constraints on the design space. The default is \code{limits = NULL}, i.e., there are no constraints. \\ \hline
\code{progress} & For \code{ace} only. A Boolean indicating whether progress of the ACE algorithm is printed. The default is \code{progress = FALSE}.\\ \hline
\code{binary} & A Boolean indicating whether the Bayesian two sample t-test (\code{FALSE}; the default) or the test of proportions (\code{TRUE}) is carried out. \\ \hline
\code{deterministic} & A logical argument indicating use of a Monte Carlo (\code{FALSE}, default) or deterministic (\code{TRUE}) approximation to the expected utility.\\ \hline
\code{mc.cores} & For \code{pace} only. The number of cores to use, i.e. at most how many child processes will be run simultaneously. The default is \code{mc.cores = 1}.\\ \hline
\code{n.assess} & For \code{pace} only. If \code{deterministic = \rev{FALSE}}, the approximate expected utility for the $C$ final designs will be calculated as the mean of \code{n.assess} approximations to the expected utility for each design. \\ \hline
\end{tabulary}
\end{center}
\caption{Optional arguments to the \code{ace} and \code{pace} functions. \label{tab:acearg2} }
\end{table}

To demonstrate the use of the \code{ace} and \code{pace} functions we use a simple Poisson response model for estimating a single parameter $\bgamma = \left(\theta\right)$. Consider an experiment where the $i$th run involves specifying the $k=1$ variable $x_i \in [-1,1]$ and measuring the count response $y_i$. Assume the following model;
$$y_i \sim \mathrm{Poisson}(\mu_i),$$
independently, for $i=1,\dots,n$, with $\mu_i = \exp(x_i \theta)$. We assume a priori that $\theta \sim \mathrm{N}\left(0,1\right)$. We find a design that maximises the expectation of the following utility function
$$
u(\theta,\by,\bd) = u(\theta, \bd) = \mathcal{I}(\theta;\,\bd)\,,
$$
where
$$\mathcal{I}(\theta;\,\bd) = \sum_{i=1}^n x_i^2 \exp\left(\theta x_i\right)$$
is the Fisher information and $\bd = \left(x_1,\dots, x_n\right)^\top$. As $u$ does not depend on $\by$, the expected utility reduces to
\begin{equation}\label{eq:aoptlog}
U(\bd) = \int u(\theta,\bd) \pi(\theta)\, \mathrm{d}\theta\,,
\end{equation}
where $\pi(\theta)$ is the density of the standard normal distribution. It is straightforward to show that
$$U(\bd) = \sum_{i=1}^n x_i^2 \exp \left( x_i^2/2  \right),$$
and the optimal design is $\bd^* = \left( \pm 1, \dots, \pm 1 \right)^\top$. However to demonstrate the use of the \code{ace} and \code{pace} functions, we employ a Monte Carlo approximation to $U(\bd)$ in Equation~\ref{eq:aoptlog}. This approximation is implemented in the \proglang{R} function below which takes two arguments: an $n \times 1$ matrix $\bd$ and the Monte Carlo sample size $B$. It returns a vector of length $B$, where each element is an evaluation of $u(\theta,\bd)$ for a value of $\theta$ generated from the prior distribution.

\begin{Schunk}
\begin{Sinput}
R> utilfisher <- function(d, B) {
+   theta <- rnorm(B)
+   ui <- matrix(rep(d[, 1] ^ 2, B), ncol = B) * exp(outer(d[, 1], theta))
+   apply(ui, 2, sum)
+   }
\end{Sinput}
\end{Schunk}

We now call the function \code{ace} to demonstrate finding a design with $n=12$ runs from a single starting design. The first mandatory argument \code{utility} is set to be the function defined above. The second mandatory argument \code{start.d} specifies the starting design; note that although $k=1$, the starting design still needs to be an \proglang{R} \code{matrix} object. Here we use a matrix of $n$ zeros. We keep all other arguments as their default values and set a seed for reproducibility.

\begin{Schunk}
\begin{Sinput}
R> set.seed(1)
R> n <- 12
R> start.d <- matrix(0, nrow = n, ncol = 1)
R> ex22a <- ace(utility = utilfisher, start.d = start.d)
\end{Sinput}
\end{Schunk}

Printing the resulting \code{"ace"} object summarises the inputs and the computing resources required.

\begin{Schunk}
\begin{Sinput}
R> ex22a
\end{Sinput}
\begin{Soutput}
User-defined model & utility 
Number of runs = 12

Number of factors = 1

Number of Phase I iterations = 20

Number of Phase II iterations = 100

Computer time = 00:00:33
\end{Soutput}
\end{Schunk}

An \code{"ace"} object is a list which includes the final designs from Phase 1 (\code{phase1.d}) and Phase 2 (\code{phase2.d}) of the algorithm. If \code{N1 = 0} (i.e., there are no Phase I iterations), then \code{phase1.d} will be equal to the argument \code{start.d}. Correspondingly, if \code{N2 = 0} (i.e., there are no Phase II iterations), then \code{phase2.d} will be equal to \code{phase1.d}. 

%


Consider now repeating the above implementation of ACE from $C=10$ different randomly generated starting designs, where each element in the design is generated uniformly from $[-1,1]$. The starting designs are organised into a list and we then call the function \code{pace} with the same \code{utility} argument as the call to \code{ace} above.
\begin{Schunk}
\begin{Sinput}
R> C <- 10
R> start.d <- list()
R> for(i in 1:C){
+ start.d[[i]] <- matrix(runif(n = n, min = -1, max = 1), ncol = 1)
+ }
R> 
R> ex22b <- pace(utility = utilfisher, start.d = start.d)
\end{Sinput}
\end{Schunk}

Printing the resulting \code{pace} object summarises the inputs and the computing resources required. 

\begin{Schunk}
\begin{Sinput}
R> ex22b
\end{Sinput}
\begin{Soutput}
User-defined model & utility 
Number of repetitions = 10

Number of runs = 12

Number of factors = 1

Number of Phase I iterations = 20

Number of Phase II iterations = 100

Computer time = 00:05:41
\end{Soutput}
\end{Schunk}

A \code{pace} object is a list which includes the Phase II design from each repetition (\code{final.d}) and, from those, the design (\code{d}) found with the largest approximate expected utility. 

To compare two designs, \code{acebayes} provides the the \code{S3} method \code{assess}. It takes two mandatory arguments: \code{d1} and \code{d2}, specifying the two designs to be compared. The argument \code{d1} should be either a \code{ace} or \code{pace} object and the two designs will be compared on the basis of the expected utility used for \code{d1}. The argument \code{d2} should either be a \code{ace}, \code{pace} or \code{matrix} object. We use \code{assess} to compare the single starting design of $n$ zeros, to the final designs from a single repetition of ACE and from $C$ repetitions. In cases like this, where the approximation to the expected utility is \rev{not deterministic}, \code{assess} will calculate \code{n.assess} approximations to the expected utility where \code{n.assess} is an optional argument. We set \code{n.assess} to be 100. The function \code{assess} will return an object of type \code{assess}. For a non-deterministic approximation to the expected utility, when printed, an \code{assess} object will show the mean and standard deviation of the \code{n.assess} evaluations of the approximate expected utility for each design.

\begin{Schunk}
\begin{Sinput}
R> assess(d1 = ex22a, d2 = matrix(rep(0, n), ncol = 1), n.assess = 100)
\end{Sinput}
\begin{Soutput}
Mean (sd) approximate expected utility of d1 = 19.78256 (0.1546731) 
Mean (sd) approximate expected utility of d2 = 0 (0) 
\end{Soutput}
\begin{Sinput}
R> assess(d1 = ex22b, d2 = ex22a, n.assess = 100)
\end{Sinput}
\begin{Soutput}
Mean (sd) approximate expected utility of d1 = 19.78025 (0.1106841) 
Mean (sd) approximate expected utility of d2 = 19.79534 (0.1633979) 
\end{Soutput}
\end{Schunk}
We can clearly see the improvement in approximate expected utility from the design of $n$ zeros and the designs found by ACE. It appears that, in this case, the $C$ repetitions of ACE have not led to any improvement in the design performance. This can be seen by plotting the \code{assess} object providing side-by-side boxplots of the \code{n.assess} evaluations of the approximate expected utility for each design.
\begin{Schunk}
\begin{Sinput}
R> assess22 <- assess(d1 = ex22a, d2 = ex22b, n.assess = 100)
R> plot(assess22)
\end{Sinput}
\begin{figure}

{\centering \includegraphics[width=4in]{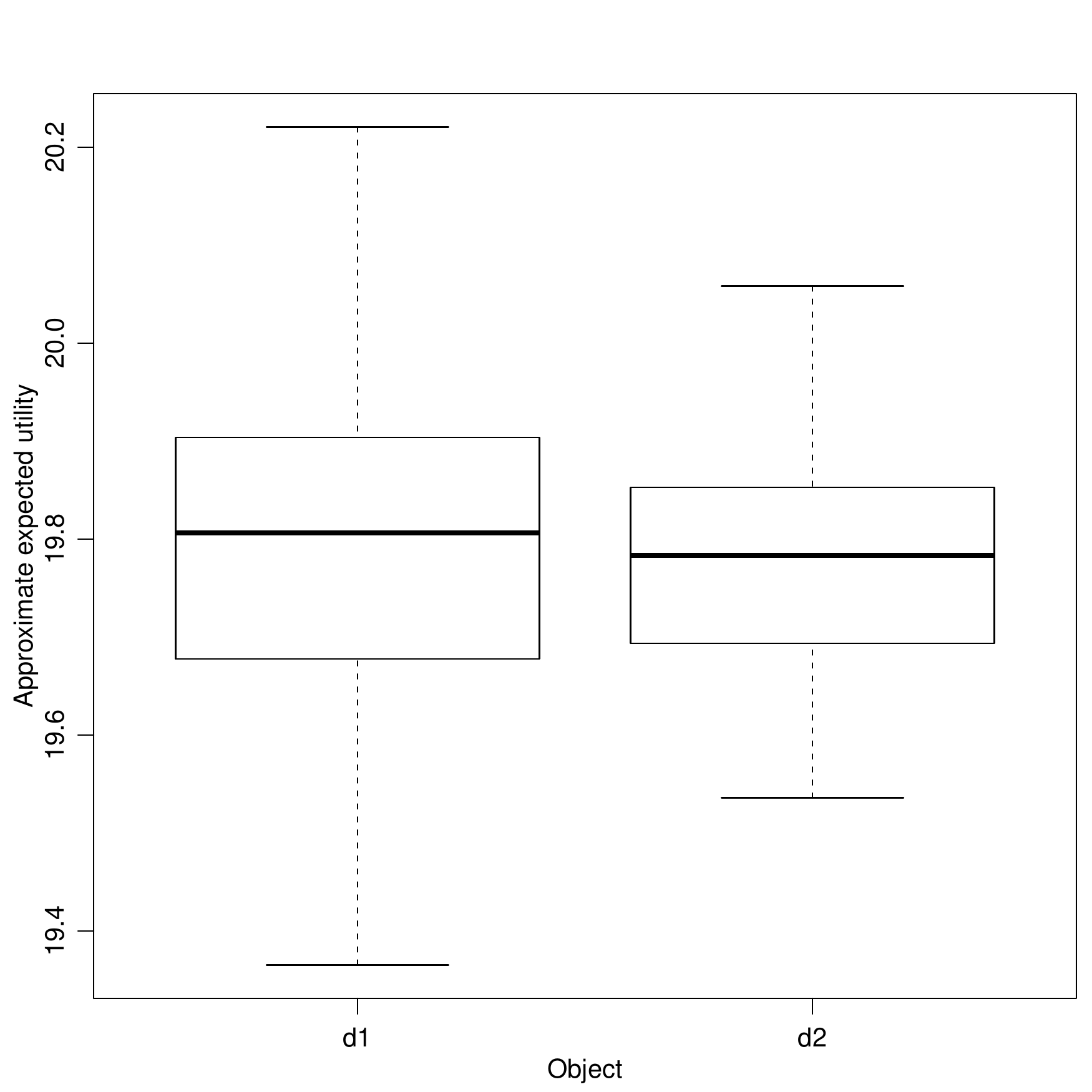} 

}

\caption[Boxplots of the evaluations of the approximate expected utility for the designs found from one (	exttt{d1}) and $C=10$ (	exttt{d2}) repetitions of ACE]{Boxplots of the evaluations of the approximate expected utility for the designs found from one (	exttt{d1}) and $C=10$ (	exttt{d2}) repetitions of ACE.}\label{fig:EX_22_8}
\end{figure}
\end{Schunk}
The resulting plot is shown in Figure~\ref{fig:EX_22_8} from which we can see that the performance of the designs are very close. This can be confirmed by inspecting the two designs and noting that they both consist only of values $x \pm 1$, i.e. both are optimal.
\begin{Schunk}
\begin{Sinput}
R> t(ex22a$phase2.d)
\end{Sinput}
\begin{Soutput}
     [,1] [,2] [,3] [,4] [,5] [,6] [,7] [,8] [,9] [,10] [,11] [,12]
[1,]   -1   -1   -1    1   -1   -1   -1   -1   -1    -1    -1    -1
\end{Soutput}
\begin{Sinput}
R> t(ex22b$d)
\end{Sinput}
\begin{Soutput}
     [,1] [,2] [,3] [,4] [,5] [,6] [,7] [,8] [,9] [,10] [,11] [,12]
[1,]   -1   -1   -1   -1   -1   -1   -1    1    1     1    -1     1
\end{Soutput}
\end{Schunk}

\section[Utilities and approximations]{Utility functions and approximations}
\label{sec:utility}

Prior to application of the ACE algorithm, a relevant, and perhaps pragmatic, choice of utility function must be made that encapsulates the aim of the experiment. In Section~\ref{sec:examples}, we illustrate use of functions from \pkg{acebayes} by finding efficient designs under three common utility functions. 

\subsection{Common utility functions}
\label{sec:cuf}

\begin{enumerate}
\item
\textbf{Shannon information gain} (SIG; \citealt{lindley_1956}):
\begin{align}
u_{SIG}(\bgamma,\by,\bd) & = \log \pi(\bgamma | \by, \bd) -  \log \pi(\bgamma | \bd) \label{eqn:sig} \\
 & = \log \pi(\by | \bgamma, \bd) -  \log \pi(\by | \bd)\,.\nonumber
\end{align}
A SIG-optimal design that maximises the expectation of $u_{SIG}$ equivalently maximises the expected Kullback-Liebler divergence between the prior and posterior distributions \citep{ChalonerVerdinelli}.
\item
\textbf{Negative squared error loss} (NSEL; e.g., \citealt{Chaloner1984}):
\begin{equation}
u_{NSEL}(\bgamma,\by,\bd) = - \left[\bgamma - \mathrm{E}(\bgamma | \by,\bd)\right]^\top\left[\bgamma - \mathrm{E}(\bgamma | \by,\bd)\right]\,.
\label{eqn:nsel}
\end{equation}
A  NSEL-optimal design that maximises the expectation of $u_{NSEL}$ equivalently minimises the expected trace of the posterior variance matrix. Note that the use of this utility is not appropriate for nominal or ordinal $\bgamma$ (for example, if $\bgamma$ holds binary model indicator variables).

\item
\textbf{0-1 utility} (e.g., \citealt{Felsenstein1992}):
\begin{equation}
u_{01}(\bgamma,\by,\bd) = \prod_{l=1}^p I\left(M_l(\by,\bd)-\delta_l < \gamma_l < M_l(\by,\bd)+ \delta_l\right)\,,
\label{eqn:01}
\end{equation}
where $M_l(\by, \bd) = \argmax_{\gamma_l}\pi(\gamma_l | \by, \bd)$ is the marginal posterior mode of $\gamma_l$, $I(A)$ is the indicator function for event $A$, and $\delta_l \ge 0$ is a specified tolerance. This utility is only non-zero if the posterior mode is ``close'' to $\gamma_l$ for all $l=1,\ldots,p$. Setting $\delta_l = 0$ is typically only appropriate for discrete $\gamma_l$.
\end{enumerate}

\subsection[Approximating utility functions]{Approximating utility functions}
\label{sec:approxutil}

Most utility functions, including those in Section~\ref{sec:cuf}, require approximation of posterior quantities, for example the marginal likelihood, $\pi(\by | \mathbf{d})$, or posterior mean, $\mathrm{E}\left(\bgamma | \by,\bd\right)$. Such quantities are analytically intractable for most models. Here we review those methods implemented in \pkg{acebayes} to produce approximate utilities $\tilde{u}(\bgamma,\by,\bd)$. 

\citet{woods_overstall_2016} used Monte Carlo approximations, with a sample $\left\{\bgamma_b\right\}_{b=1}^{\tilde{B}}$ from $\pi(\bgamma | \bd)$, to approximate $u_{SIG}$ and $u_{NSEL}$ (in Equations~\ref{eqn:sig} and~\ref{eqn:nsel}, respectively) to enable design selection for parameter estimation, i.e., with $\bgamma = \btheta$, for a single model. When combined with a Monte Carlo approximation to the expected utility with sample size $B$, the resulting \textbf{nested Monte Carlo} (or double-loop Monte Carlo) approximation to $U(\bd)$ is subject to bias of order $\tilde{B}^{-1}$ (see \citealp{Ryan2003}). Hence, large values of both $B$ and $\tilde{B}$ are required to achieve suitable precision for design comparison and neglible bias, resulting in computationally expensive utility approximations.



Although the use of the one-dimensional emulators $\hat{U}_{ij}(x)$ helps to alleviate the computational cost associated with a nested Monte Carlo approximation, the adoption of alternative, cheaper, utility approximations can further increase the range and size of design problems that can be addressed. Several classes of analytical approximations have been proposed using normal approximations to the posterior distribution. \citet{overstall_etal_2016} applied ACE with a normal approximation to the posterior distribution with mean equal to the posterior mode and variance-covariance matrix equal to the inverse of the expected Fisher information, $\mathcal{I}(\btheta;\,\bd)$, minus the second derivative of the log prior density, both evaluated at the posterior mode. Such an approximation can lead to analytically tractable, if still potentially biased, \textbf{normal-based} approximations $\tilde{u}(\bgamma, \by, \bd)$; for example, via a Laplace approximation to the marginal likelihood (see also \citealp{Laplace}).   


Simpler approximations to some utilities can be obtained by using $\mathcal{I}(\btheta;\,\bd)^{-1}$ as an approximation to the posterior variance-covariance matrix \citep[e.g.,][]{ChalonerVerdinelli}. For example, for estimation of $\bgamma = \boldsymbol{\theta}$, approximations to the SIG and NSEL utility functions are given by
\begin{align}
\tilde{u}_{SIGD}(\boldsymbol{\theta},\by,\bd) & =  \log |\mathcal{I}(\boldsymbol{\theta};\, \bd)|\,, \label{eqn:Dopt}\\
\tilde{u}_{NSELA}(\boldsymbol{\theta},\by,\bd) & =  - \mathrm{tr} \left\{\mathcal{I}(\boldsymbol{\theta};\,\bd)^{-1}\right\}\,. \label{eqn:Aopt}
\end{align}
Designs that maximise the expectation of $\tilde{u}_{SIGD}$ and $\tilde{u}_{NSELA}$ with respect to the prior distribution of $\btheta$ are referred to as \textbf{pseudo-Bayesian} $D$- and $A$-optimal, respectively. 


\subsection{Approximating the expected utility}
\label{sec:approxexputil}

The \pkg{acebayes} package uses expected utility approximations of the form given in Equation~\ref{eqn:approxEU}. For the nested Monte Carlo and normal-based approximations, $U(\bd)$ is approximated by the sample mean of $\tilde{u}(\bgamma_b,\by_b,\bd)$ for a sample $\left\{\bgamma_b,\by_b\right\}_{b=1}^B$ from $\pi(\bgamma,\by | \bd)$. Default values of $B$ (and $\tilde{B}$ for nested Monte Carlo) are $B=1000$ for constructing the one dimensional emulators $\hat{U}_{ij}(x)$ and $B=20,000$ when calculating the probability of accepting the proposed design (see Section~\ref{sec:ace}).

For pseudo-Bayesian $D$- and $A$-optimal design, where the approximations given by Equations~\ref{eqn:Dopt} and~\ref{eqn:Aopt} do not depend on $\by$, the $p$-dimensional integrals with respect to $\bgamma=\btheta$ can be approximated using quadrature methods. The \pkg{acebayes} package implements a radial-spherical integration rule \citep{MG1997}, with $\bgamma_b$ and $\omega_b$ in Equation~\ref{eqn:approxEU} being abscissas and (non-constant) weights, respectively. For small $p$, the value of $B$ is typically of the order of several hundred, making this approach much less computationally intensive than either nested Monte Carlo or normal-based methods. Both multivariate normal and independent uniform prior densities are implemented for use with quadrature approximations in \pkg{acebayes}.  See \citet{gotwalt_jones_steinberg_2009} for more details on using this quadrature scheme to find pseudo-Bayesian $D$-optimal designs.   

\section[Examples]{Examples}
\label{sec:examples}

In this section, we demonstrate the use of the \pkg{acebayes} package to find Bayesian optimal designs for four examples. Despite ACE being able to find efficient designs for larger and more complex problems than existing methods in the literature, it still requires significant computational resources. Hence we have chosen examples that illustrate the main features of the methodology and package but that do not require excessive computer time to complete. It should be clear how the examples can be extended to address more complex or realistic scenarios. In particular, the main arguments to the functions \code{ace} and \code{pace} are essentially identical. Therefore, to minimise the computer time taken to reproduce the examples, we only demonstrate the \code{pace} functionality in Sections~\ref{sec:comp} and~\ref{sec:logreg} (in addition to that already demonstrated in Section~\ref{sec:acebayes}).

A particular feature of this section is demonstration of the functions \rev{\code{aceglm}} and \rev{\code{acenlm}}, which simplify the process of finding Bayesian optimal designs for generalised linear models and nonlinear models, respectively. The \code{ace} function allows designs to be sought for very general problems. This flexibility comes at the price that non-expert users may feel uncomfortable with the level of additional coding required to use the function. To remove this potential barrier to the use of the package, \code{aceglm} and \code{acenlm} provide wrappers to \code{ace} that implement the ACE algorithm for these common model types. In both cases, the functions allow designs to be found for parameter estimation under a single model for a range of (approximated) utility functions. There are also corresponding functions, \code{paceglm} and \code{pacenlm}, which implement repetitions of ACE for generalised linear models and nonlinear models, respectively. Both \code{(p)aceglm} and \code{(p)acenlm} make use of the familiar \code{formula} and \code{family} \proglang{R} arguments and objects. In Sections~\ref{sec:comp} and ~\ref{sec:logreg} we demonstrate, in detail, the use of these functions.


In each case, unless otherwise stated, we use a randomly generated Latin hypercube space-filling design \citep[see][ch.~5]{SWN2003} as the starting design for the ACE algorithm. These designs are generated using the \code{randomLHS} function in the \pkg{lhs} package \citep{lhs_pack}. Each element of the starting design is scaled to lie in the stated design space. Additionally, before we generate such a design, we set a random seed for full reproducibility of the results in this section.

\subsection[Compartmental non-linear model]{Compartmental non-linear model}
\label{sec:comp}

In this example, we demonstrate using \code{acenlm} to generate a pseudo-Bayesian $D$-optimal design for a compartmental model commonly used in pharmacokinetics (PK). The \code{acenlm} and \code{pacenlm} functions can find optimal designs for models of the form 
\begin{equation}
y_i \sim \mathrm{N}\left(\mu(\btheta;\,\bx_i), \sigma^2\right)\,,
\label{eqn:nlm}
\end{equation}
where $y_1,\ldots, y_n$ are assumed independently distributed and the user specifies a non-linear function, $\mu(\btheta;\,\bx_i)$, of parameters $\btheta$, and prior distributions for both $\btheta$ and $\sigma^2 >0$, the unknown response variance. 

A PK experiment typically involves introducing a fixed amount of drug to the body at time zero and measuring at times $t_1,\ldots,t_n$ the amount of drug remaining in the body. Hence the design consists of the $n$ sampling times, i.e., $\bx_i = (t_i)$, $k=1$ and $\bd = (t_1,\ldots,t_n)^\top$. Here we assume a sampling interval such that $t_i\in[0, 24]$ hours. We illustrate \code{(p)acenlm} on the compartmental model
$$\mu(\btheta;\,t_i) = \theta_3 \left[ \exp \left(-\theta_1 t_i\right) - \exp \left(-\theta_2 t_i\right) \right]\,,$$
with $p=3$ unknown parameters $\btheta = \left(\theta_1,\theta_2,\theta_3\right)^\top$. 

For non-linear models of the form of Equation~\ref{eqn:nlm}, the Fisher information for $\boldsymbol{\theta}$ is
$$\mathcal{I}(\boldsymbol{\theta};\,\bd) = \frac{1}{\sigma^2} \sum_{i=1}^n \frac{ \partial \mu(\boldsymbol{\theta};\,t_i) }{\partial \boldsymbol{\theta}}\frac{ \partial \mu(\boldsymbol{\theta};\,t_i) }{\partial \boldsymbol{\theta}^\top}.$$

Following \citet{gotwalt_jones_steinberg_2009}, we find designs with $n=18$ sampling times and assume that elements of $\btheta$ have the following independent prior distributions: 
\begin{equation}
\theta_1 \sim \mathrm{U} \left[0.01884,0.9884\right]\,,\quad \theta_2 \sim \mathrm{U} \left[0.298,8.298\right]\,,
\label{eqn:nlprior}
\end{equation}
with $\theta_3$ having a prior point mass at 21.8. 

To find a pseudo-Bayesian $D$-optimal design using \code{acenlm} we first specify one starting design having a single column named \code{"t"} and with elements scaled to the sampling interval $[0,24]$.

\begin{Schunk}
\begin{Sinput}
R> set.seed(1)
R> n <- 18
R> k <- 1
R> p <- 3
R> start.d <- randomLHS(n = n, k = k) * 24
R> colnames(start.d) <- c("t")
\end{Sinput}
\end{Schunk}

We use quadrature to approximate the expected utility, and hence define the \code{prior} below as a list containing a single matrix \code{support}, where the rows specify the lower and upper limits of the uniform prior distribution for each parameter, respectively; see Equation~\ref{eqn:nlprior}. For use with the \code{acenlm} function, the columns of this matrix should be named. As $\mathcal{I}(\boldsymbol{\theta};\,\bd)$ only depends linearly on $1/\sigma^2$, the relative performance of designs under pseudo-Bayesian criteria do not depend on the unknown $\sigma^2$. Hence it is not required to specify a prior distribution for $\sigma^2$.

\begin{Schunk}
\begin{Sinput}
R> a1 <- c(0.01884, 0.298)
R> a2 <- c(0.09884, 8.298)
R> prior <- list(support = cbind(rbind(a1, a2), c(21.8, 21.8)))
R> colnames(prior[[1]]) <- c("theta1", "theta2", "theta3") 
\end{Sinput}
\end{Schunk}

The \code{acenlm} function takes three mandatory arguments. The argument \code{formula} gives a symbolic description of the non-linear model (in this example, \code{~ theta3 * (exp( - theta1 * t) - exp( - theta2 * t))}). The argument \code{start.d} specifies the starting design, an $n \times k$ matrix with columns named as per the terms in the \code{formula} argument; and \code{prior} specifies the prior distribution for $\btheta$, with the input for this argument depending on the chosen \code{method} (see below). In addition, we set the lower and upper limits of each element of the design space to be 0 and 24 respectively. 

\begin{Schunk}
\begin{Sinput}
R> ex411 <- acenlm(formula = ~ theta3 * (exp( - theta1 * t) - 
+   exp( - theta2 * t)), start.d = start.d, prior = prior, lower = 0, 
+   upper = 24)
\end{Sinput}
\end{Schunk}
For demonstration purposes, we compare the designs found from Phase I and II using the \code{S3} method \code{assess} (as introduced in Section~\ref{sec:acebayes}). In cases of a deterministic approximation to the expected utility, \code{assess} will calculate one approximation to the expected utility for each of \code{d1} and \code{d2}. Furthermore, when \code{d1} is as a result of a call to \code{acenlm} or \code{aceglm} with the \code{criterion} argument being \code{"D"} or \code{"A"}, then \code{assess} will also \rev{approximate, again using quadrature,} the pseudo-Bayesian relative $D$- or $A$-efficiency. The pseudo-Bayesian relative $D$-efficiency of design $\bd_1$ relative to design $\bd_2$, defined as
$$
\mbox{Deff}(\bd_1, \bd_2) = 100 \times \exp\left\{\left[U_D(\bd_1) - U_D(\bd_2)\right] / p\right\}\,, 
$$
provides a quantitative comparison of two designs. Here, 
$$
U_D(\bd) = \int \log|\mathcal{I}(\btheta;\,\bd)|\,\pi(\btheta) \,\mathrm{d}\btheta\,.
$$
Similar relative efficiency exists for pseudo-Bayesian $A$-optimal designs. \rev{The function \code{assess} calculates these efficiencies for pseudo-Bayesian $D$ and $A$-optimal designs.}
\begin{Schunk}
\begin{Sinput}
R> assess(d1 = ex411, d2 = ex411$phase1.d)
\end{Sinput}
\begin{Soutput}
Approximate expected utility of d1 = 15.79695 
Approximate expected utility of d2 = 15.70753 
Approximate relative D-efficiency = 103.0255% 
\end{Soutput}
\end{Schunk}
\rev{Here} we see that Phase II led to a small increase in the expected utility. We can also compare the two designs in terms of the number of unique sampling times.
\begin{Schunk}
\begin{Sinput}
R> length(unique(ex411$phase1.d))
\end{Sinput}
\begin{Soutput}
[1] 18
\end{Soutput}
\begin{Sinput}
R> length(unique(ex411$phase2.d))
\end{Sinput}
\begin{Soutput}
[1] 13
\end{Soutput}
\end{Schunk}
Therefore Phase II consolidated the design into 13 unique sampling times.

It is quite common in PK, and similar, experiments for there to be constraints on the minimum time between successive measurements. For such experiments, the ordered sampling times, $t_{(1)},\dots,t_{(n)}$, should satisfy the following constraint:
\begin{equation}
\min_{i=1,\dots,n-1} |t_{(i)} - t_{(i+1)}| > c\,.
\label{eqn:compcon}
\end{equation}  
See \citet{ryan_drovandi_thompson_pettitt_2014} and \citet{woods_overstall_2016} for examples with similar constraints on the design for the compartmental model.

We can include such constraints in Phase I of the ACE algorithm using the \code{limits} argument. In Phase I, the candidate design is chosen by replacing the current coordinate (i.e., $x_{ij}$, $i=1,\ldots,n;\,j=1,\ldots,k$) by the value that maximises the predictive mean of the Gaussian process emulator. As discussed in Section~\ref{sec:ace}, this maximisation is achieved by choosing the point with largest predicted mean from a grid of points, typically on an interval defined by the arguments \code{lower} and \code{upper}. We can also specify the argument \code{limits} as a function to incorporate (multivariate and dynamic) constraints on the design coordinates. The function should have three arguments: \code{d}, \code{i} and \code{j}. Here \code{d} specifies the design, and \code{i} and \code{j} define the current coordinate by specifying the row and column of \code{d}. The function should return a grid of points (in the form of a vector) from which the point with the largest predicted mean will be chosen. 

The code below defines a \code{limits} function to incorporate the constraint given by Equation~\ref{eqn:compcon}, with $c=0.25$ (i.e., 15 minute intervals between sampling). A grid of 10,000 points from \code{lower = 0} to \code{upper = 24} is created. We then remove all points from this grid that are within 15 minutes of the sampling times in the current design excluding the $i$th point. Note that here the function does not depend on \code{j} as the design involves only $k=1$ factor.

\begin{Schunk}
\begin{Sinput}
R> limits <- function(d, i, j) {
+   grid<-seq(from = 0, to = 24, length.out = 10000)
+   for(s in as.vector(d)[-i]) {
+     grid <- grid[(grid < (s - 0.25)) | (grid > (s + 0.25))]
+   }
+   grid
+ }
\end{Sinput}
\end{Schunk}

We find a design satisfying the constraint by including the specification of the \code{limits} argument in the \code{acenlm} function and setting the number of Phase II iterations to $N_2=0$ (as we do not want to consolidate clusters into repeated sampling times; see the constraint given by Equation~\ref{eqn:compcon}).

\begin{Schunk}
\begin{Sinput}
R> ex412a <- acenlm(formula = ~ theta3 * (exp( - theta1 * t) - 
+   exp( - theta2 * t)), start.d = start.d, prior = prior, lower = 0, 
+   upper = 24, limits = limits, N2 = 0)
\end{Sinput}
\end{Schunk}

As described in Section~\ref{sec:ace}, we should actually repeat ACE from $C$ different starting designs. This can be achieved using the \code{pacenlm} function which takes the same arguments as \code{acenlm}. The only exception is that the argument \code{start.d} should be a list where each element is an $n \times k$ matrix. Below we specify such a list with $C=10$.
\begin{Schunk}
\begin{Sinput}
R> C <- 10
R> start.d <- list()
R> for(i in 1:C){
+ start.d[[i]] <- randomLHS(n = n, k = k) * 24
+ colnames(start.d[[i]]) <- c("t")
+ }
R> ex412b <- pacenlm(formula = ~ theta3 * (exp( - theta1 * t) - 
+   exp( - theta2 * t)), start.d = start.d, prior = prior, lower = 0, 
+   upper = 24, limits = limits, N2 = 0)
\end{Sinput}
\end{Schunk}
We compare approximations to the expected utility for the design found from one repetition of ACE against the design found under $C=10$ repetitions. 
\begin{Schunk}
\begin{Sinput}
R> assess(d1 = ex412a, d2 = ex412b)
\end{Sinput}
\begin{Soutput}
Approximate expected utility of d1 = 15.34813 
Approximate expected utility of d2 = 15.36236 
Approximate relative D-efficiency = 99.52679% 
\end{Soutput}
\end{Schunk}
The design found under one repetition (\code{ex412a$phase2.d}) is about 99.5$\%$ $D$-efficient compared to the design found under $C=10$ repetitions. In this case, the procedure was relatively robust to the starting design.

\subsection[Logistic regression]{Logistic regression}
\label{sec:logreg}

In this section we consider a logistic regression model from \citet{woods_overstall_2016} to demonstrate the use of the \code{aceglm} and \code{paceglm} functions. We find designs that maximise the expected NSEL utility for estimation of parameters $\bgamma = \btheta$ using two different approximations to the utility function: 1) approximation given by Equation~\ref{eqn:Aopt}, resulting in a pseudo-Bayesian $A$-optimal design; and 2) the normal-based approximation of \citet{overstall_etal_2016}.

A binary response is assumed to depend on $k=4$ variables through the model
$$y_i \sim \mathrm{Bernoulli}(\rho_i)\,,$$
for $i=1,\ldots,n$ with $\rho_i = 1/[1+\exp(-\eta_i)]$,
$$\eta_i = \theta_0 + \sum_{j=1}^4 \theta_j x_{ij}\,,$$
and $\boldsymbol{\theta} = \left(\theta_0,\theta_1,\theta_2,\theta_2,\theta_4\right)^\top$ being the $p=5$ unknown parameters that require estimation. We find designs with $n=6$ runs, $\bx_{i} = (x_{i1},\ldots,x_{i4})^\top$, assuming $-1\le x_{ij}\le 1$ ($i=1,\ldots,6; j=1,\ldots, 4$). The dimension of the design space is $nk = 24$.

Independent uniform prior distributions for each element of $\boldsymbol{\theta}$ are assumed,
\begin{equation}\label{eq:logisticprior}
\begin{array}{llllllll}
\theta_0 \sim \mathrm{U}[-3,3]\,, & & \theta_1\sim\mathrm{U}[4,10]\,, & &\theta_2\sim\mathrm{U}[5,11]\,,\\
\theta_3\sim \mathrm{U}[-6,0]\,, & & \theta_4 \sim \mathrm{U}[-2.5,3.5]\,.\\
\end{array}
\end{equation}


The \code{aceglm} function has four mandatory arguments. The arguments \code{formula} and \code{family} are the well-known arguments we would supply to the \code{glm} function in the \pkg{stats} package \citep{stats_pack} and are used to define the logistic regression model, i.e., \code{formula = ~ x1 + x2 + x3 + x4} and \code{family = binomial}. The argument \code{start.d} specifies the starting design, an $n \times k$ matrix with columns named as per the terms in the \code{formula} argument; and \code{prior} specifies the prior distribution for $\btheta$, with the input for this argument depending on the chosen \code{method} (see below). Initially, we specify values for $n$, $p$ and $k$, and generate a list of $C=10$ starting designs.

\begin{Schunk}
\begin{Sinput}
R> set.seed(1)
R> n <- 6
R> p <- 5
R> k <- 4
R> C <- 10
R> start.d <- list()
R> for(i in 1:C){
+ start.d[[i]] <- randomLHS(n = n, k = k) * 2 - 1
+ colnames(start.d[[i]]) <- c("x1", "x2", "x3", "x4")
+ }
\end{Sinput}
\end{Schunk}

\subsubsection[Pseudo-Bayesian $A$-optimal design]{Pseudo-Bayesian $A$-optimal design}


We start by finding a pseudo-Bayesian $A$-optimal design by specifying \code{criterion = "A"}. Note that the default value is \code{criterion = "D"} which would result in a pseudo-Bayesian D-optimal design using the approximation given in Equation~\ref{eqn:Dopt}. To approximate the expected utility we use the radial-spherical quadrature rule discussed in Section~\ref{sec:approxexputil}. This can be specified by setting the \code{method} argument in \code{paceglm} to \code{"quadrature"}, which is the default for pseudo-Bayesian criteria (i.e., \code{criterion} being \code{"A"}, \code{"D"} or \code{"E"}). Under this \code{method}, to specify the uniform prior distribution given by Equation~\ref{eq:logisticprior}, we define a list with a single matrix argument \code{support} with the first row specifying the lower limits for the prior for each parameter, and the second row specifying the upper limits. The other optional arguments to \code{paceglm} are shared with \code{ace} (see Table~\ref{tab:acearg2}). The exception is the argument \code{deterministic}, which is not required for \code{(p)aceglm} and \code{(p)acenlm} as the \code{method} argument specifies if the approximation to the expected utility is deterministic. We leave these optional arguments set to default values but demonstrate their use in later examples. 

%

\begin{Schunk}
\begin{Sinput}
R> a1 <- c(-3, 4, 5, -6, -2.5)
R> a2 <- c(3, 10, 11, 0, 3.5)
R> prior <- list(support = rbind(a1, a2))
R> ex411 <- paceglm(formula = ~ x1 + x2 + x3 + x4, family = binomial, 
+   start.d = start.d, prior = prior, criterion = "A")
\end{Sinput}
\end{Schunk}

%
%

For demonstration purposes, using the \code{assess} function, we compare the design found with the highest expected utility from $C=10$ repetitions (i.e. \code{ex411$d}) and the design found under the first repetition (i.e. \code{ex411$final.d[[1]]}).

\begin{Schunk}
\begin{Sinput}
R> assess(d1 = ex411, d2 = ex411$final.d[[1]])
\end{Sinput}
\begin{Soutput}
Approximate expected utility of d1 = -225.6464 
Approximate expected utility of d2 = -267.3872 
Approximate relative A-efficiency = 118.4983% 
\end{Soutput}
\end{Schunk}

This clearly demonstrates the advantage of repeating ACE from different starting designs.

\subsubsection[Normal-based approximation to the NSEL utility function]{Normal-based approximation to the NSEL utility function}

To apply the normal-based approximation to the NSEL utility function we let \code{criterion = "NSEL-Norm"} in the \code{paceglm} function. This changes the default \code{method} argument to \code{"MC"} implementing Monte Carlo. Under this method, we require an \proglang{R} function that generates a sample from the prior distribution given by Equation~\ref{eq:logisticprior}. 

\begin{Schunk}
\begin{Sinput}
R> prior <- function(B) {
+   theta <- matrix(0, nrow = B, ncol = p)
+   for(b in 1:B) {
+     theta[b, ] <- runif(n = p, min = a1, max = a2)
+     }
+   theta
+ }
\end{Sinput}
\end{Schunk}

Again, we set all other optional arguments to their default values.

\begin{Schunk}
\begin{Sinput}
R> ex412 <- paceglm(formula = ~ x1 + x2 + x3 + x4, family = binomial, 
+   start.d = start.d, prior = prior, criterion = "NSEL-Norm")
\end{Sinput}
\end{Schunk}

We can check the approximate convergence of the ACE algorithm using the \code{S3} method \code{plot.ace}.

\begin{Schunk}
\begin{Sinput}
R> plot(ex412)
\end{Sinput}
\begin{figure}

{\centering \includegraphics[width=4in]{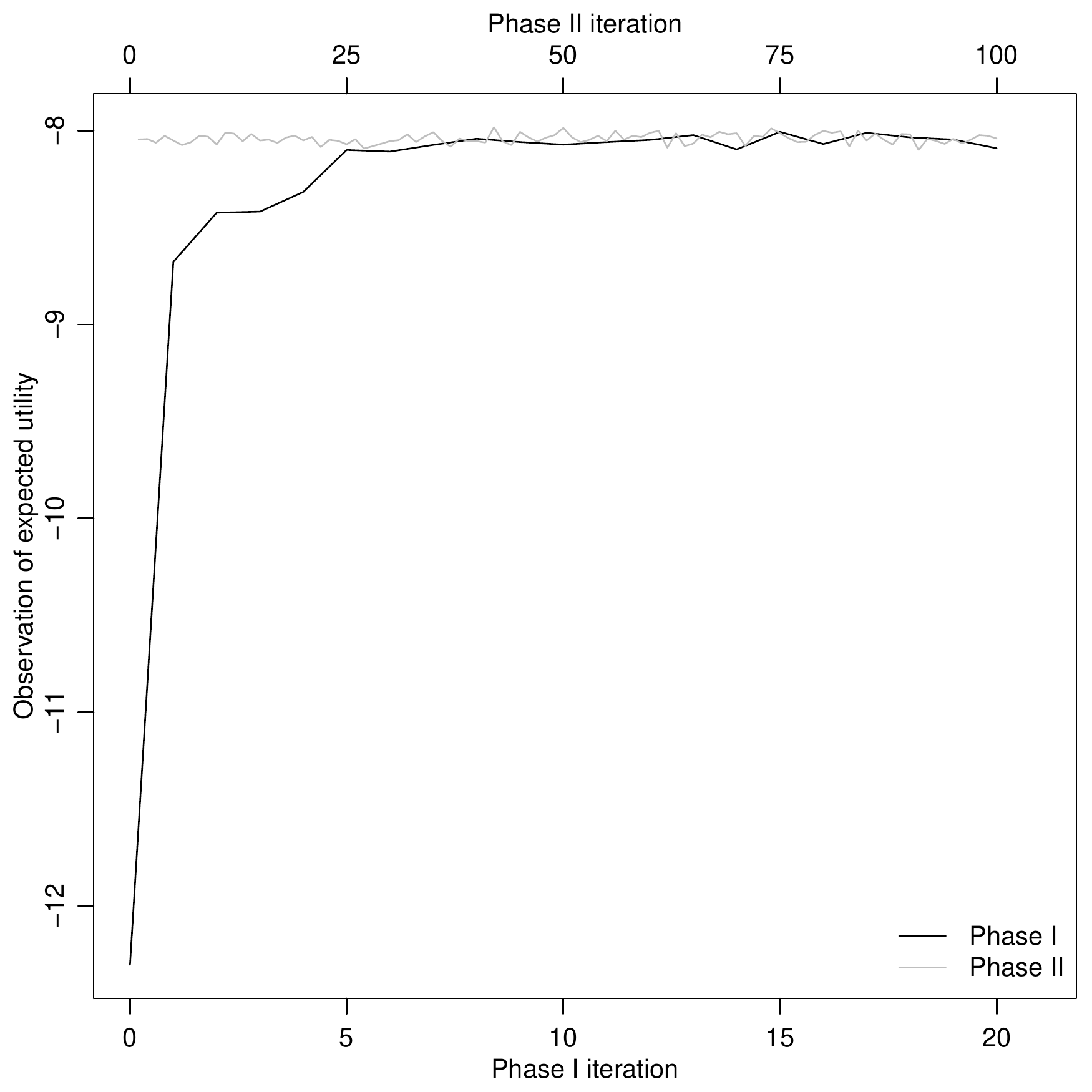} 

}

\caption[Convergence of phases I (lower $x$-axis) and II (upper $x$-axis) for the logistic regression model under the normal-based approximation to the NSEL utility function]{Convergence of phases I (lower $x$-axis) and II (upper $x$-axis) for the logistic regression model under the normal-based approximation to the NSEL utility function.}\label{fig:EX_42_4}
\end{figure}
\end{Schunk}

This function produces a trace plot (see Figure~\ref{fig:EX_42_4}) of the Monte Carlo approximation to the expected utility~\eqref{eqn:exputil} against iteration number for Phases I and II and the design that had the highest expected utility from the $C$ repetitions. In Phase I, the algorithm makes very large initial improvements to the approximate expected utility, and appears to have converged after six or seven iterations. Phase II does not appear to lead to any improvements in the design, as also occurred when finding the pseudo-Bayesian $A$-optimal design.


We now compare the pseudo-Bayesian $A$-optimal design (\code{ex411$d}) to the design found under the normal-based approximation to the NSEL utility (\code{ex412$d}). 




\begin{Schunk}
\begin{Sinput}
R> assess(d1 = ex412, d2 = ex411)
\end{Sinput}
\begin{Soutput}
Mean (sd) approximate expected utility of d1 = -8.056569 (0.0367072) 
Mean (sd) approximate expected utility of d2 = -10.61123 (0.03650658) 
\end{Soutput}
\end{Schunk}

Notice how the optimal design under the normal-based approximation (\code{ex412$d}) achieves a substantially larger expected NSEL utility than the pseudo-Bayesian $A$-optimal design (\code{ex411$d}). \citet{woods_overstall_2016} and \citet{overstall_etal_2016} empirically investigated how the difference between NSEL and $A$-optimal designs decreases as $n$, the number of runs, increases, as the asymptotic approximation underpinning pseudo-Bayesian optimal designs improves. \citet{overstall_etal_2016} also found that, for this example, the difference between designs found using ACE with nested Monte Carlo and normal-based approximations to the expected NSEL utility were negligible, regardless of the value of $n$. However, designs under the normal-based approximation typically took around one-third of the computational time to find.

\subsection[Model selection for chemical reactions]{Model selection for chemical reactions}

In this example, we demonstrate using the \code{ace} function for a problem that falls outside the capabilities of \code{aceglm} and \code{acenlm}, and illustrate construction of a bespoke \code{utility} function.

This example is adapted from \citet{box_hill_1967} and concerns mechanistic modelling of chemical reactions. We find designs with $n$ runs where the $i$th run requires specifying the reaction time, $x_{i1} \in (0,150)$, and temperature, $x_{i2} \in (450,600)$, at which to measure reaction yield, $y_i$ ($i=1,\ldots,n$). Therefore each run of the design is a two-vector $\bx_{i} = (x_{i1}, x_{i2})^\top$. The following statistical model is posited:
$$y_i \sim \mathrm{N}\left(\mu(m,\boldsymbol{\theta};\,\bx_i),\sigma^2\right)\,,$$
where
$$
\mu(m, \boldsymbol{\theta};\bx_i) = 
\left\{
\begin{array}{cc}
\exp \left( - \eta(\boldsymbol{\theta};\,\bx_i)\right) & \mbox{for } m = 0 \\ \\
\left[1 + m\eta(\btheta;\,\bx_i)\right]^{-\frac{1}{m}} & \mbox{for } m = 1,2,3\,.
\end{array}
\right.
$$
Here
$$\eta(\boldsymbol{\theta};\,\bx_i) = \theta_1 x_{i1} \exp \left( - \frac{\theta_2}{x_{i2}}\right)\,,$$
with unknown parameters $\boldsymbol{\theta} = \left(\theta_1,\theta_2\right)^\top$ and $m\in\mathcal{M}=\{0,1,2,3\}$ specifying the order of the reaction, with $m=0,1,2,3$ corresponding to first-, second-, third- and fourth-order reactions, respectively.

We create an \proglang{R} function to implement $\eta(\btheta;\,\bx_i)$ with arguments \code{d}, an $n \times k$ matrix, and \code{theta}, a $B \times p$ matrix with $b$th row given by $\boldsymbol{\theta}_b = (\theta_{1b}, \theta_{2b})^\top$. It returns a $B \times n$ matrix with $bi$th element given by $\theta_{1b}x_{i1}\exp \left( -  \theta_{2b}x_{i2} \right)$. That is, it calculates the value of the function $\eta$ for a given design $\bd$ for every row of a matrix of parameter values \code{theta}.

\begin{Schunk}
\begin{Sinput}
R> etafunc <- function(d, theta) {
+   outer(theta[, 1], d[, 1]) * exp( - outer(theta[, 2], 1 / d[, 2]))
+ }
\end{Sinput}
\end{Schunk}


The aim of the experiment is to determine which order reaction is appropriate for the observed responses, i.e., a choice from the set $\mathcal{M} = \left\{0,1,2,3\right\}$. That is, model selection with $\bgamma = (m)$ in $u(\bgamma,\by,\bd)$. Following \citet{overstall_etal_2016}, identical prior distributions are assumed for the parameters under each model:
\begin{equation}\label{eqn:chemprior}
\theta_{1} \sim \mathrm{N}\left(400,25^2\right)\,, \quad \theta_{2}  \sim \mathrm{N}\left(5000,250^2\right)\,.
\end{equation}
We fix the response standard deviation as $\sigma=0.1$, double the value assumed by \citet{box_hill_1967}. We choose this larger value as use of too small a value of $\sigma$ leads to the expected utility becoming less dependent on the design, i.e., the expected utility surface is quite flat. Equal prior probabilities are assumed for each model, i.e., $\pi(m) = 1/4$, for all $m \in \mathcal{M}$.

\begin{Schunk}
\begin{Sinput}
R> sig <- 0.1
R> prior <- function(B) {
+   theta1 <- rnorm(n = B, mean = 400, sd = 25)
+   theta2 <- rnorm(n = B, mean = 5000, sd = 250)
+   cbind(theta1, theta2)
+ }
\end{Sinput}
\end{Schunk}

We aim to find a design that maximises the expected 0-1 utility, i.e., the expectation of $u_{01}$ given by Equation~\ref{eqn:01}, and set $\delta_1 = \delta = 0$. For equal prior model probabilities, the posterior modal model, $M(\by,\bd)$, will maximise the marginal likelihood given by $\pi(\by | m,\bd) = \int\pi(\by | m,\bd,\btheta)\pi(\btheta)\,\mathrm{d}\btheta$. As the marginal likelihood is not available in closed form for these models, we use a Monte Carlo approximation implemented as a \code{utility} function which can then be passed to the \code{ace} function to find an optimal design.

We approximate $u_{01}$ by
\begin{equation}
\tilde{u}_{01}(m,\by,\bd) = I(m = \tilde{M}(\by,\bd))\,,
\end{equation}
where
\begin{align*}
\tilde{M}(\by,\bd) & = \argmax~\tilde{\pi}(\by | m,\bd)\,, \\
\tilde{\pi}(\by | m,\bd) & = \frac{1}{\tilde{B}}\sum_{b=1}^{\tilde{B}}\pi(\by | m,\tilde{\btheta}_b,\bd)\,, 
\end{align*}
with $\{\tilde{\btheta}_b\}_{b=1}^{\tilde{B}}$ a sample from the prior distribution given by Equation~\ref{eqn:chemprior}, likelihood $\pi(\by | m, \btheta, \bd) = \prod_{i=1}^n\pi(y_i | m, \btheta, \bx_i)$ and $\pi(y_i | m, \btheta, \bx_i)$ being a normal density with mean $\mu(m, \btheta;\,\bx_i)$ and variance $\sigma^2$.

The expected 0-1 utility can then be approximated as
$$
\tilde{U}(\bd) = \frac{1}{B}\sum_{b=1}^B\tilde{u}(m_b,\by_b,\bd)\,,
$$
for a sample $\{m_b,\btheta_b,\by_b\}_{b=1}^B$ from the joint distribution of $m$, $\btheta$ and $\by$; such a sample can be easily generated by sampling a model indicator from $\mathcal{M}$ with probabilities $\pi(m)$, parameters from the prior distribution with density $\pi(\btheta)$, and then responses from the conditional distribution with density $\pi(\by | m, \btheta)$ (see the code below).

We can now create an \proglang{R} function, \code{util01}, to implement this approximate utility. Note that we set $\tilde{B}=100$. The function takes as arguments a design \code{d} and Monte Carlo sample size \code{B}. The main work is done within the nested \code{for} loop; for each data set generated in the outer loop, the posterior modal model is found by maximising a Monte Carlo approximation to the marginal likelihood, which is calculated in the inner loop. It returns a vector of \code{B} evaluations of the Monte Carlo approximated utility function $\tilde{u}(m,\by,\bd)$.

\begin{Schunk}
\begin{Sinput}
R> Btilde <- 100
R> 
R> util01 <- function(d, B) {
+ 
+   theta <- prior(B)
+   mod <- sample(x = 0:3, size = B, replace = TRUE)
+ 
+   eta <- etafunc(d = d, theta = theta)
+   mu <- matrix(0, nrow = B, ncol = n)
+   mu[mod == 0, ] <- exp( - eta[mod == 0, ])
+   mu[mod == 1, ] <- (1 + eta[mod == 1, ]) ^ ( - 1)
+   mu[mod == 2, ] <- (1 + 2 * eta[mod == 2, ]) ^ ( - 1 / 2)
+   mu[mod == 3, ] <- (1 + 3 * eta[mod == 3, ]) ^ ( - 1 / 3)
+ 
+   Y <- mu + sig * matrix(rnorm(B * n), nrow = B)
+ 
+   thetatilde <- prior(Btilde)
+   etatilde <- etafunc(d = d, theta = thetatilde)
+   mutilde0 <- exp(-etatilde)
+   mutilde1 <- (1 + etatilde) ^ ( - 1)
+   mutilde2 <- (1 + 2 * etatilde) ^ ( - 1 / 2)
+   mutilde3 <- (1 + 3 * etatilde) ^ ( - 1 / 3)
+ 
+   modal <- rep(0, B)
+   for(b in 1:B) {
+     C <- matrix(0, nrow = Btilde, ncol = 4)
+     for(bt in 1:Btilde) {
+       C[bt, 1] <- sum(dnorm(x = Y[b, ], mean = mutilde0[bt, ], 
+       sd = sig, log = TRUE))
+       C[bt, 2] <- sum(dnorm(x = Y[b, ], mean = mutilde1[bt, ], 
+       sd = sig, log = TRUE))
+       C[bt, 3] <- sum(dnorm(x = Y[b, ], mean = mutilde2[bt, ], 
+       sd = sig, log = TRUE))
+       C[bt, 4] <- sum(dnorm(x = Y[b, ], mean = mutilde3[bt, ], 
+       sd = sig, log = TRUE))
+     }
+     logmarglik <- log(apply(exp(C), 2, mean))
+     modal[b] <- which.max(logmarglik) - 1
+   }
+   ifelse(modal == mod, 1, 0)
+ }
\end{Sinput}
\end{Schunk}

We use the \code{ace} function along with \code{util01} to find an optimal design. For illustration, we search for a design with $n=20$ runs and set the arguments \code{B = c(1000, 100)}, \code{Q = 15}, \code{N2 = 0} (to skip Phase II), and using the arguments \code{lower} and \code{upper} to set the bounds for each factor (see Table~\ref{tab:acearg2}). We specify a Bayesian hypothesis test of a difference in proportions by setting \code{binary = TRUE}.

\begin{Schunk}
\begin{Sinput}
R> set.seed(1)
R> n <- 20
R> q <- 2
R> lower <- cbind(rep(0, n), rep(450, n))
R> upper <- cbind(rep(150, n), rep(600, n))
R> start.d <- randomLHS(n = n, k = q)*(upper - lower) + lower
R> ex43 <- ace(utility = util01, start.d = start.d, B = c(1000, 100), Q = 15, 
+   N2 = 0, binary = TRUE, lower = lower, upper = upper)
\end{Sinput}
\end{Schunk}

Note that $C$ repetitions of the ACE algorithm could have been implemented using the \code{pace} function specifying a list of starting designs as the argument \code{start.d}. However we have only demonstrated \code{ace} here to minimise the computing time required to reproduce the example.

We compare the approximations to the expected 0-1 utility for the starting design and the final design from ACE as follows.



\begin{Schunk}
\begin{Sinput}
R> assess43 <- assess(d1 = ex43, d2 = start.d)
R> assess43
\end{Sinput}
\begin{Soutput}
Mean (sd) approximate expected utility of d1 = 0.8789 (0.00611211) 
Mean (sd) approximate expected utility of d2 = 0.80565 (0.0120973) 
\end{Soutput}
\end{Schunk}

Assuming the data generating process is consistent with one of the models, the starting design identifies the true model about 81\% of the time. By using an optimal design we can increase this to about 88\%.

\subsection[Prediction]{Optimal design for prediction}
Prediction from a nonparametric regression model is a common aim of both spatial studies and computer experiments, often using Gaussian process (GP) regression (or Kriging). For example, optimal sensor placements, e.g., for pollution or environmental monitoring, may be sought within a geographical region of interest to provide accurate predictions at unsampled locations. See \citet{Diggle_Lophaven}, \citet{Zimmerman} and \citet{Ucinski}. As the cost of maintaining large monitoring networks can be high, costs are often also associated with the placement of each sensor. We demonstrate using \pkg{acebayes} to find an optimal design for such a problem.   

Here, a design consists of $n$ locations $\bx_i = (x_{i1}, x_{i2})^\top$, within a specified two-dimensional region, at each of which a response $y_i$ will be observed. The aim is to fit a Gaussian process model to the resulting responses, $\by = (y_1,\ldots,y_n)^\top$, and to predict the $n_0$ unobserved responses, $\by_0 = (y_{01},\ldots,y_{0n_0})^\top$, where $y_{0i}$ is associated with pre-specified location $\bx_{0i} =  (x_{0i1}, x_{0i2})^\top$, for $i=1,\dots,n_0$. Let $\tilde{\by} = \left(\by^\top,\by_0^\top\right)^\top$ denote the $\tilde{n} \times 1$ vector of observed and unobserved responses where $\tilde{n} = n + n_0$. 

The design problem is to specify the $n \times 2$ matrix $\bd$ (with $i$th row $\bx_i^\top$) to best predict $\by_0$ where the meaning of \textquotedblleft best" is controlled by the choice of utility function (see later).

A zero-mean Gaussian process model results in the assumption of multivariate normal distribution for $\tilde{\by}$,
\begin{equation}
\tilde{\by} |\sigma^2, \phi, \tau^2 \sim N\left(\mathbf{0}, \sigma^2 \tilde{\bSigma}\right)\,,
 \label{eqn:GP}
\end{equation}
where $\sigma^2>0$ is a scale parameter and 
\begin{equation}
\tilde{\bSigma}=\tilde{\mathbf{C}}+\tau^2 \mathbf{I}_{\tilde{n}}.
 \label{eqn:C}
\end{equation}
In Equation~\ref{eqn:C}, $\mathbf{I}_{\tilde{n}}$ is the $\tilde{n} \times \tilde{n}$ identity matrix, $\tau^2>0$ is the nugget, and $\tilde{\mathbf{C}}$ is an $\tilde{n} \times \tilde{n}$ correlation matrix partitioned as follows
\begin{equation}
\tilde{\mathbf{C}} = \left( \begin{array}{cc}
\mathbf{C} & \mathbf{S} \\
\mathbf{S}^\top &  \mathbf{C}_0 \end{array} \right).
 \label{eqn:part}
\end{equation}
In Equation~\ref{eqn:part}, $\mathbf{C}$ is an $n \times n$ matrix with $rt$th element
$$C_{rt} = \rho(\bx_r,\bx_t, \phi), \quad \mbox{for $r,t = 1,\dots,n$,}$$
$\mathbf{C}_0$ is an $n_0 \times n_0$ matrix with $rt$th element
$$C_{0rt} = \rho(\bx_{0r},\bx_{0t}, \phi), \qquad \mbox{for $r,t = 1,\dots,n_0$,}$$
$\mathbf{S}$ is an $n \times n_0$ matrix with $rt$th element
$$S_{rt} = \rho(\bx_{r},\bx_{0t}, \phi), \qquad \mbox{for $r = 1,\dots,n$ and $t=1,\dots,n_0$,}$$
and $\rho$ is a known correlation function. In this example, we employ the squared exponential correlation function, $\rho(\bx_k,\bx_l;\phi) = \exp\{-\phi\sum_{j=1}^2(x_{kj}-x_{lj})^2\}$, as implemented by the following \proglang{R} function.

\begin{Schunk}
\begin{Sinput}
R> rho <- function(X1, X2, phi) {
+   k <- ncol(X1)
+   n1 <- nrow(X1)
+   n2 <- nrow(X2)
+   A <- matrix(0, nrow = n1, ncol = n2)
+   for(i in 1:k) {
+     A <- A - phi * (matrix(rep(X1[, i], n2), nrow =n1) - 
+     matrix(rep(X2[, i], each = n1), nrow = n1)) ^ 2
+   }
+   exp(A)
+ }
\end{Sinput}
\end{Schunk}

We adopt a Bayesian approach with conjugate prior distribution assigned to the parameter $\sigma^2$: 
$$
\sigma^{-2}\sim \mbox{Gamma}\left(\frac{a}{2},\frac{b}{2}\right)\,,
$$
where $a=3$ and $b=1$ are known shape and rate parameters, respectively. The correlation parameter $\phi=0.5$ and nugget $\tau^2=1\times 10^{-5}$ are assumed known and fixed. 

\begin{Schunk}
\begin{Sinput}
R> phi <- 1
R> tau2 <- 0.00001
R> a <- 3
R> b <- 1
\end{Sinput}
\end{Schunk}

We assume that the two-dimensional region is such that $0 \le x_{ij} \le 1$. The $n_0$ pre-specified locations $\bx_{01},\dots,\bx_{0n_0}$ are given by the points on an evenly-spaced $r \times r$ grid where $r=10$, i.e., $n_0 = 100$. We let $\bd_0$ be the $n_0 \times 2$ matrix with $i$th row given by $\bx_{0i}^\top$, for $i=1,\dots,n_0$.

\begin{Schunk}
\begin{Sinput}
R> k <- 2
R> r <- 10
R> n0 <- r ^ k
R> x0 <- seq(from = 0, to = 1, length.out = r)
R> d0 <- as.matrix(expand.grid(x0, x0))
\end{Sinput}
\end{Schunk}


We find optimal designs for prediction using a utility function adapted from \citet{SansoMuller1997} and \citet{Muller2004} which compromises between the accuracy of the posterior predictive mean, $\mathrm{E}(\by_0 | \by)$, at the $n_0$ new locations $\bx_{01},\ldots,\bx_{0n_0}$ against the cost of the $n$ placed sensors:
\begin{equation}\label{eqn:util_pred}
u(\by_0,\by,\bd) = \sum_{i=1}^{n_0} I(\mathrm{E}(y_{0i} | \by) - \delta < y_{0i} < \mathrm{E}(y_{0i} | \by) + \delta) - \sum_{i=1}^n c(\bx_i) \,,
\end{equation}
where $\delta>0$ controls the desired accuracy and $c(\bx_i)$ is the cost of taking an observation at $\bx_i$. In this example, the cost $c(\bx_i)$ depends on the location of the proposed sensor and is given by
$c(\bx_i) = x_{i1}^2 + x_{i2}^2$, i.e., the squared Euclidean distance from the origin. The total cost of the design is given by $\sum_{i=1}^n c(\bx_i)$. Under the model given by Equation~\ref{eqn:GP}, the posterior predictive mean of $\by_0$ is given by
\begin{equation}
\mathrm{E}(\by_0|\by)= \mathbf{S}^\top \left(\mathbf{C} + \tau^2 \mathbf{I}_n \right)^{-1}\by\,,
\end{equation}
with $i$th element $\mathrm{E}(y_{0i} | \by)$, for $i=1,\dots,n_0$.


A Monte Carlo approximation to the expectation of $u(\by_0,\by,\bd)$ given by Equation~\ref{eqn:util_pred} can be constructed as
$$
\tilde{U}(\bd) = \frac{1}{B}\sum_{b=1}^B u(\by_{0b}, \by_{b}, \bd)\,,
$$
with $\tilde{\by}_b = (\by_b^\top,\by_{0b}^\top)^\top$ sampled from the marginal distribution of $\tilde{\mathbf{y}}$. 

The function below implements this approximation, returning a vector of \code{B} evaluations of $u(\by_0,\by,\bd)$. We have specified $\delta = 0.25$.

\begin{Schunk}
\begin{Sinput}
R> delta <- 0.25
R> 
R> utilpred <- function(d, B) {
+ 
+   n <- dim(d)[1]
+   C <- rho(d, d, phi)
+   S <- rho(d, d0, phi)
+   C0 <- rho(d0, d0, phi)
+   
+   sig2 <- 1 / rgamma(n = B, shape = 0.5 * a, rate = 0.5 * b)
+   Sigmatilde <- rbind(cbind(C, S), cbind(t(S), C0)) + tau2 * diag(n + n0)
+   cSigmatilde <- chol(Sigmatilde)
+   ytilde <- matrix(0, nrow = n + n0, ncol = B)
+   for(b in 1:B) {
+     ytilde[, b] <- sqrt(sig2[b]) * t(rnorm(n + n0) %*% cSigmatilde)
+   }
+   
+   y <- ytilde[1:n, ]
+   y0 <- ytilde[ - (1:n), ]
+   
+   postpredmean <- t(S) %*% solve(C + tau2 * diag(n)) %*% y
+   
+   accuracy <- rep(0,B)
+   for(b in 1:B) {
+     accuracy[b] <- sum(as.numeric(((postpredmean[, b] - delta) < y0[, b]) * 
+       ((postpredmean[, b] + delta) > y0[,b])))
+   }
+   
+   cost <- sum(apply(d^2, 1, sum))
+   
+   accuracy - cost
+ }
\end{Sinput}
\end{Schunk}

We now illustrate the use of \code{ace} with these functions by finding a design with $n=10$ sensors. 

\begin{Schunk}
\begin{Sinput}
R> set.seed(1)
R> n <- 10
R> start.d <- randomLHS(n = n, k = k)
R> ex44 <- ace(utility = utilpred, start.d = start.d, lower = 0, upper = 1)
\end{Sinput}
\end{Schunk}


We can compare the value of the objective function for the starting design and the optimal design obtained from ACE.


\begin{Schunk}
\begin{Sinput}
R> assess(d1 = ex44, d2 = start.d)
\end{Sinput}
\begin{Soutput}
Mean (sd) approximate expected utility of d1 = 95.86214 (0.02456676) 
Mean (sd) approximate expected utility of d2 = 92.41169 (0.02434313) 
\end{Soutput}
\end{Schunk}

We also compare the cost of each design.

\begin{Schunk}
\begin{Sinput}
R> sum(apply(start.d ^ 2, 1, sum))
\end{Sinput}
\begin{Soutput}
[1] 6.653732
\end{Soutput}
\begin{Sinput}
R> sum(apply(ex44$phase2.d ^ 2, 1, sum))
\end{Sinput}
\begin{Soutput}
[1] 3.081133
\end{Soutput}
\end{Schunk}

\begin{Schunk}
\begin{figure}

{\centering \includegraphics[width=4in]{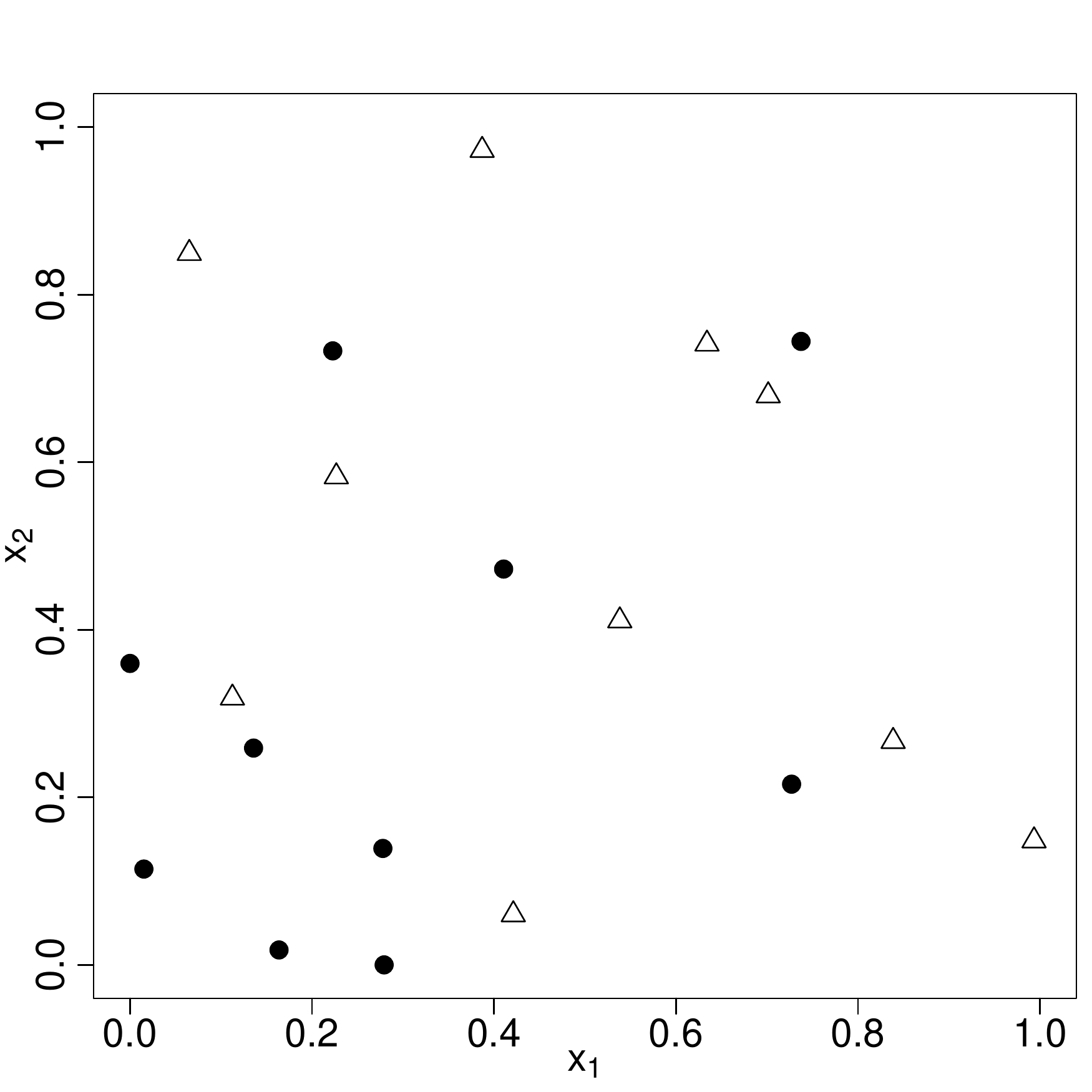} 

}

\caption[Bayesian optimal ($\bullet$) and starting ($\triangle$) designs for the prediction example]{Bayesian optimal ($\bullet$) and starting ($\triangle$) designs for the prediction example.}\label{fig:EX_44_8}
\end{figure}
\end{Schunk}

The difference in utility values between the Bayesian optimal design and the starting design are mostly due to the much lower cost of the sensor placements in the optimal design. That is, the optimal design has similar predictive accuracy as the starting design but at substantially reduced cost. Figure~\ref{fig:EX_44_8} shows the very different distribution of the points in the optimal design compared to the starting design, with the former having many more points in areas of low cost near the origin. 

\section[Discussion]{Discussion}
\label{sec:disc}

Bayesian optimal design is conceptually straightforward but often difficult, and computationally expensive, to implement. The \pkg{acebayes} package provides a suite of functions that allow optimal designs to be found for complex and realistic problems, with dimensionality at least one order of magnitude greater than other current methods. The general purpose \code{ace} and \code{pace} functions can be used to solve very general, and bespoke, design problems. The functions \code{(p)aceglm} and \code{(p)acenlm} can find designs for common classes of statistical models.

Any set of examples can only be illustrative, and those in this paper are no different. To aid exposition, we have deliberately kept the problems relatively simple and the designs sought have been small. However, \pkg{acebayes} has been used to find designs for more complex scenarios. These include high-dimensional design spaces \citep{woods_overstall_2016} and ordinary differential equation models \citep{overstall_odes_2016}. It has also been used \citep{overstall_mcgree_2018} to find design for intractable likelihood models \citep[e.g.][]{DrovandiPettitt} and for synthesis optimisation of pharmaceutical products \citep{overstall_etal_2018}. 

While \pkg{acebayes} allows much larger designs to be found than existing methods, for complex problems we would recommend coding the utility function in a low-level programming language (e.g., \proglang{C/C++}) and running the code on a computational cluster. The algorithm is heuristic, and so to overcome convergence to local optima, it should be run multiple times from different starting designs, for example using parallel computing \rev{and/or the \code{pace} function}.

We have created a YouTube channel (\url{https://bit.ly/2SSC3ur}) hosting tutorial videos and a Twitter account (\href{https://twitter.com/acebayes}{\code{@acebayes}}) to update users on the latest developments to the \pkg{acebayes} package.

\section*{Acknowledgments}
D.C. Woods was partially supported by a Fellowship EP/J018317/1 from the UK Engineering and Physical Sciences Research Council. The authors are grateful for constructive comments from two anonymous reviewers that improved the paper. 

\appendix

\section{The approximate coordinate exchange algorithm} \label{app:ace}

This appendix provides details on Phase I (Appendix~\ref{alg:ace1}) and Phase II (Appendix~\ref{alg:ace2}) of the ACE algorithm.

\subsection{Phase I} \label{alg:ace1}
\begin{enumerate}
\item \label{alg:start}
Choose an initial design $\bd^0 \in \mathcal{D}$ and set the current design to be $\bd^C = \bd^0$.
\item
For $i=1,\dots,n$ and $j=1,\dots,k$, complete the following steps.
		\begin{enumerate}
		\item
		Let $\bd^C(x_{ij}^q)$ equal $\bd^C$ with $ij$th coordinate (entry) replaced by $x_{ij}^q$, where $x_{ij}^1, \ldots, x_{ij}^Q$ are the points from a one-dimensional space filling design in $\mathcal{D}_{ij}$, the design space for the $ij$th element of $\bd$.
    \item \label{alg:approx}
    For $q=1,\ldots,Q$, evaluate $\tilde{U}[\bd^C(x_{ij}^q)]$, the approximation to the expected utility, e.g., Equation~\ref{eqn:approxEU}. Fit a Gaussian process emulator to ``data'' $\left\{x_{ij}^q, \tilde{U}[\bd(x_{ij}^q)]\right\}_{q=1}^Q$, and set $\hat{U}_{ij}(x)$ to be the resulting predictive mean.
		\item
		Find
		$$x_{ij}^\star = \argmax_{x \in \mathcal{D}_{ij}} \hat{U}_{ij}(x)\,,$$
		and set $\bd^\star = \bd^C(x_{ij}^\star)$.
		\item \label{alg:1_2e}
		For a stochastic (e.g., Monte Carlo) approximation $\tilde{U}$, set $\bd^C = \bd^\star$ with probability $p^*$ derived from a Bayesian hypothesis test. For a deterministic approximation (e.g., quadrature), set $\bd^C=\bd^\star$ if $\tilde{U}(\bd^\star)>\tilde{U}(\bd^C)$. 
		\end{enumerate}
\item
Repeat Step 2 $N_1$ times.
\end{enumerate}

\subsection{Phase II} \label{alg:ace2}
\begin{enumerate}
\item
Set the current design, $\bd^C$, to be the final design from Phase I of the ACE algorithm.
\item \label{alg:1}
For $i=1,\dots,n$, set
$$\bd_i^{(1)} = \left[(\bd^C)^\top,(\bx^C_i)^\top\right]^\top\,,$$
where $(\bx_i^C)^\top$ is the $i$th row of $\bd^C$; that is, form $\bd^{(1)}_i$ by augmenting $\bd^C$ with a repeat of the $i$th run.
\item
Find $i^\star = \argmax_{i=1,\dots,n} \tilde{U}(\bd_i^{(1)})$ and set $\bd^{(2)} = \bd_{i^\star}^{(1)}$.
\item
For $h=1,\dots,n+1$, set
$$\bd^{(3)}_h = \left[(\bx_{1}^{(2)})^\top,\ldots,(\bx_{h-1}^{(2)})^\top, (\bx_{h+1}^{(2)})^\top, \ldots, (\bx_{n+1}^{(2)})^\top\right]^\top\,,$$
where $(\bx_h^{(2)})^\top$ is the $h$th row of $\bd^{(2)}$; that is, form $\bd^{(3)}_h$ by removing the $h$th run. 
\item
Find $h^\star = \argmax_{h=1,\dots,n+1} \tilde{U}(\bd^{(3)}_h)$ and set $\bd^* = \bd_{h^\star}^{(3)}$.
\item \label{alg:2_2e}
For a stochastic (e.g., Monte Carlo) approximation $\tilde{U}$, set $\bd^C = \bd^\star$ with probability $p^*$ derived from a Bayesian hypothesis test. For a deterministic approximation (e.g., quadrature), set $\bd^C=\bd^\star$ if $\tilde{U}(\bd^\star)>\tilde{U}(\bd^C)$. 
\item 
Repeat steps \ref{alg:1} to \ref{alg:2_2e} $N_2$ times.
\end{enumerate}

\bibliography{acebayes}

\end{document}